\documentclass[a4paper,aps,pra,showpacs,preprintnumbers]{revtex4}
\usepackage{graphics,graphicx,epsfig}
\usepackage[centertags]{amsmath}
\usepackage{amsfonts}
\usepackage{amssymb}
\usepackage[english]{babel}
\usepackage{graphicx}
\usepackage{amsthm,color}

\begin{document}

\def\BE{\begin{equation}}
\def\EE{\end{equation}}

\def\BY{\begin{eqnarray}}
\def\EY{\end{eqnarray}}

\def\BM{\begin{multline}}
\def\EM{\end{multline}}

\def\L{\label}
\def\nn{\nonumber}
\def\({\left (}
\def\){\right)}
\def\<{\langle }
\def\>{\rangle}
\def\[{\left [}
\def\]{\right]}
\def\o{\overline}
\def\BA{\begin{array}}
\def\EA{\end{array}}
\def\ds{\displaystyle}
\def\dsp{\displaystyle}

\title{Quantum correlations and fluctuations in the pulsed light produced by a synchronously pumped optical parametric
oscillator below its oscillation threshold}

\author{ V.A. Averchenko, Yu.M. Golubev }
\affiliation{V.~A.~Fock Physics Institute, St.~Petersburg State University \\
198504 St.~Petersburg, Stary Petershof,
ul. Ul'yanovskaya, 1, Russia}
\author{C. Fabre, N. Treps}
\affiliation{Laboratoire Kastler Brossel, Universit\'{e} Pierre et Marie Curie-Paris6,
 Place Jussieu, CC74, 75252 Paris
Cedex 05, France}
\date{\today}

\begin{abstract}
We  present a simple quantum theory for the pulsed light generated
by a synchronously pumped optical parametric oscillator (SPOPO) in
the degenerate case where the signal and idler trains of pulses
coincide, below threshold and neglecting all dispersion effects. Our
main goal is to precise in the obtained quantum effects, which ones
are identical to the c.w. case and which ones are specific to the
SPOPO. We  demonstrate in particular that the temporal correlations
have interesting peculiarities: the quantum fluctuations at
different times within the same pulse turn out to be  totally not
correlated, whereas they are correlated between nearby pulses at
times that are placed in the same position relative to the center of
the pulses. The number of
significantly correlated pulses is of the order of cavity finesse.
We show also that there is perfect squeezing at noise frequencies
multiple of the pulse repetition frequency when one approaches the
 threshold from below on the signal field quadrature measured by a balanced homodyne detection
with a local oscillator of very short duration compared to the SPOPO
pulse length.

\end{abstract}

\maketitle


\section{Introduction \L{I}}
Optical parametric oscillators (OPOs) are well-known and efficient
sources of non-classical light. Whereas c.w. OPOs have been
extensively studied both by theoreticians and experimentalists
\cite{Bachor}, the quantum properties of pulsed OPOs have been so
far much less investigated, in spite of their interest
\cite{Slusher}. In particular, synchronously pumped OPOs (SPOPOs),
in which the pump pulses are temporally separated by the round trip
time of the OPO cavity, seem particularly promising, as theoretically shown in\cite{Valcarcel, Menicucci}. In these devices,
the efficiency in twin-photon generation is enhanced twice, because
of the high peak power in light pulses and because of field
enhancement in a resonant cavity. In addition, the light emitted by SPOPOs has been theoretically shown to be either
multi-mode squeezed or multipartite entangled, which makes it an
interesting resource for parallel transfer and processing of quantum
information \cite{Leuchs} and for quantum metrology in the time
domain, for example to measure ultra-short time delays with  a very
high sensitivity \cite{Lamine}.

A detailed quantum analysis of the SPOPO has been performed in
Refs.~\cite{Valcarcel, Patera}. In these papers the intracavity field
is considered in the frequency domain, i.e. expanded as a combination of longitudinal modes of the SPOPO cavity. It was demonstrated  that a SPOPO emits a tensor product of squeezed "super-modes", each being a particular coherent superpositions of longitudinal modes of different frequencies. In this article we will take a rather different approach of the same problem, based on a description of the device in the time domain instead of the frequency domain.  It is well-known that a theoretical description of pulsed processes is often more pictorial in such a frame, and the results obtained are complementary from the ones deduced from the frequency approach. More precisely, we will use the two-time technique used in Ref.~\cite{Haus} for the pulse laser generation.

The paper is organized as follows. In Sec.~\ref{II}  we present a physical model of the SPOPO in terms of two coupled Heisenberg-Langevin equations that describe in the time-domain the SPOPO operation both below and above oscillation threshold.
In Sec.~\ref{III} we determine the SPOPO oscillation threshold. In Sec.~\ref{IV} we construct the pair correlation functions for the signal quadrature components outside the cavity in the below threshold configuration. In Sec.~\ref{V} the current correlation functions and their spectra are considered with and without time averaging by the detection process. Some interesting quantum features of the parametric oscillation are considered in the spectral domain in Sec.~\ref{VI} and discussed in Sec.~\ref{VII}. In the Appendices a way to construct the main equations is considered in detail and typical physical parameters are selected.

\section{ Intracavity two-time description of field pulses\L{II}}
Fig. 1 presents a sketch of a SPOPO that we consider here. A $\chi^{(2)}$ nonlinear parametric crystal is inserted in an
optical ring cavity and is pumped by a train of laser pulses with mean frequency $\omega_p$. We assume that the duration of the pump pulses $\tau_p$  is much shorter than the cavity round trip time $T_R$ (under real conditions $\tau_p$ is about
$10^{-5} T_{R}$). In the crystal takes place a degenerate type I  parametric conversion of the pump field into a collinearly propagating signal field with mean frequency $\omega_p/2$, as well as the reverse process. We assume a degenerate parametric interaction for the carrier frequencies of the pump and signal modes, meaning that the following phase matching condition is fulfilled:
\BY
\Delta  k = k_p(\omega_{p})-2
k_s(\omega_{p}/2) =0
\EY
where $k_p$ and $k_s$ are wave vectors of
carriers of pump and signal fields.

The crystal is supposed to be thin so that we can neglect the influence of mismatch and dispersion of the group
velocities on parametric interaction. Changes of fields caused by the parametric interaction are considered as perturbations of free propagation inside the crystal. In
Appendix~\ref{estimations} we consider the physical parameters of the different parametric crystals and demonstrate that, for thin enough crystals, realistic conditions exist where we can avoid the undesirable influence of the dispersion phenomenon.

We suppose that the SPOPO ring cavity is resonant and of high-finesse both for pump and signal fields, so that the doubly-resonant
configuration is realized. We assume also that the cavity is dispersion-compensated by intracavity dispersive
elements. This implies that optical pulses of arbitrary shapes are not distorted after one round trip inside the cavity, and also that it takes the same time $T_R$ for pump and signal  pulses to make a single round trip inside the cavity. In addition, the device is synchronously pumped, meaning that this time is equal to the pump repetition rate: $T_R = T_p$. These hypotheses are set for simplicity, and the equations that we use in this paper could  be easily modified in order to take into account deviations from these ideal conditions.

In the degenerate parametric generation configuration that we consider here the field operator inside the cavity is equal to:
\BY
&& \hat E(z,t)=\hat E_p(z,t)+\hat E_s(z,t).\L{1}
\EY
We use the plane wave approximation, so that the field amplitudes depend only on one longitudinal coordinate $z$ measured along optical axis of the cavity. As the pump of the parametric crystal is realized by a train of optical pulses of duration close to $100 fs$ it is possible to disjoint in the standard form quick oscillations of fields with optical frequencies $\omega_{p,s}$  from slow changes of their envelopes \cite{Kolobov_99}. At
the crystal entrance for the pump $p$ and signal $s$ waves the two fields read
\BY
&& \hat E_r(z,t)=i \left(\frac{\hbar \omega_r}{2 n_r \varepsilon_0 c S} \right)^{1/2} e^{-i\omega_r
t+ik_rz}\hat A_r(z,t),\qquad r=s,p,\qquad(\omega_p=2\omega_s),\L{2}
\EY
where $n_r = n_r(\omega_r)$ are the indices of refraction of the
crystal and $k_r=k_r (\omega_r)$ are the pump and signal wave
vectors determined by linear dispersion of the crystal. The slow
amplitudes $\hat{A_r}$ are normalized so that the mean
values $\langle\hat{A_r}^{\dag} \hat{A_r}\rangle$ have the meaning of mean fluxes in photons per second for the light beams of area S. One supposes that the crystal is placed just after the coupling mirror and that the field at the input
boundary of the crystal (at $z=0$), taking into account the periodic structure of the fields, is
\BY
&& \hat A_{r}(0,t)=\sum_{n}\hat A_{r,n}(t-nT_R).
\EY
Here ${\hat A}_{r,n}(t-nT_R)$ is the envelope of the $n$-th pulse. When $t-nT_R$ appears as argument of the envelope
we will treat it as time deviation from center of the pulse, i.e., it changes in the interval $[-T_R/2,+T_R/2]$.

In order to describe SPOPO operation inside the cavity we use the
two-time approach applied by Haus in Ref.~\cite{Haus} for developing
a quantum theory of actively mode-locked lasers. We assume that the
envelopes of pulses are not significantly changed from one pulse to
the next, a hypothesis that is typically valid in experiments with
high-finesse cavity and weak parametric amplification. Then the
dependence on discrete number $n$ could be replaced approximately by
a continuously varying  temporal parameter $T$ in the following way
\BY
&&\hat A_{r,n}(t-nT_R)\to\hat A_{r}(t,T).\L{6}
\EY
In the Appendix~\ref{HLE} we show in detail how  to derive Heisenberg-Langevin equations for these envelopes. We show that in the thin crystal approximation and when one neglects the dispersion phenomenon these equations read
\begin{eqnarray}
&& \frac{\partial \hat{A}_p(t,T)}{\partial T} = - \kappa_p \(\hat{A}_p(t,T) -{\cal A}_0(t)\)
- g {\hat{A}_s^{\; 2}(t,T)} + \hat{F}_{p}(t,T), \L{HL_p}\\
&& \frac{\partial \hat{A_s}(t,T)}{\partial T} = - \kappa_s \hat{A_s}(t,T) + 2g { \hat{A}_p(t,T) }
\hat{A}_s^{\dag}(t,T) + \hat{F}_{s}(t,T) \L{HL_s}
\end{eqnarray}
Here $\kappa_{p}$ and $\kappa_{s}$ are the loss rates of pump and signal fields respectively; $g$ is
a constant characterizing the parametric coupling; ${\cal A}_0(t)$ is the classical steady-state
envelope of the pump pulses inside the cavity, depending only on time $t$ since the pump pulses are supposed to be perfectly identical.

The Langevin noise sources $\hat{F}_{p}(t,T)$ and $\hat{F}_{s}(t,T)$  are characterized by the nonzero pair correlation functions:
\begin{eqnarray}
&& \langle \hat{F}_{r}(t,T) \hat{F}_{r}^{ \dag}(t',T') \rangle = 2\kappa_r \; \delta(T-T')
 \delta(t-t'), \qquad r=s,p. \L{noise}
\end{eqnarray}

\section{Oscillation threshold of the SPOPO\L{III}}
In order to describe the parametric effect below threshold, we can neglect pump depletion, and consider only the equation~(\ref{HL_s}) for the signal field, while the pump field remains in the coherent state with c-number amplitude ${\cal A}_0(t)$ imposed by the external pumping of the SPOPO. According to our model there is no phase modulation of the pump field, therefore without loss of generality we will treat the envelope of the pump pulse as a real positive quantity ${\cal A}_0={\cal
A}^\ast_0>0$. Let us rewrite Eq.~(\ref{HL_s}) using  the quadrature components of signal field
\BY
&&\hat X_s(t,T)=\frac{1}{2}\(\hat A_s(t,T)+\hat A_s^\dag(t,T)\),\qquad\hat
Y_s(t,T)=\frac{1}{2i}\(\hat A_s(t,T)-\hat A_s^\dag(t,T)\).
\EY
One then obtains two independent equations for the $X_s$- and the $Y_s$-quadratures
 \BY
 && \frac{\partial }{\partial T}\left(\begin{array}{c}\hat X_s(t,T)\\\hat Y_s(t,T)\end{array}\right)=
- \kappa_\mp(t)\left(\begin{array}{c}\hat X_s(t,T)\\\hat Y_s(t,T)\end{array}\right)  +
\left(\begin{array}{c}\hat{F}^\prime_{s}(t, T)\\\hat{F}^{\prime\prime}_{s}(t, T)\end{array}\right)
,\L{9}
\EY
where
\begin{eqnarray}
&& \hat{F}_{s}(t,T)=\hat{F}^\prime_{s}(t,T) +i
\hat{F}^{\prime\prime}_{s}(t,T),\qquad
\kappa_\pm=\kappa_s\(1\pm\mu(t)\),\qquad
\mu(t)=\frac{2g}{\kappa_s}{\cal A}_0(t).
\end{eqnarray}
The coefficients $\kappa_-$ and $\kappa_+$  are here time dependent,
and can be negative in some time interval, leading to a divergence
of the field quadratures, i.e. to a bifurcation of the system to the
generation of "bright" pulses. The requirement $\kappa_-=0$
determines therefore the threshold of the parametric oscillation. In
order to stay below threshold we require:
\BY
&&N_0(t)={\cal A}_{0}^2(t)<N_{th}=\frac{\kappa_s^2}{4g^2}.
\EY
It is important to note that for the SPOPO the threshold depends not
on the mean power of the pump as for the continuous generation but
on the peak power of the pulse, because of the instantaneous
character of parametric interaction. This is interesting for
experimentalists, because the corresponding mean power at threshold
can be extremely small:
\BY
&&\o N=\frac{1}{T_R}\int_{-T_R/2}^{+T_R/2}dt N_0(t)\ll N_{th}
\EY

\section{Below threshold operation: correlation functions for the field quadratures outside the cavity\L{IV}}

Let us now determine the quantum fluctuations and correlations of the output signal field quadratures, defined by:
\BY
&&\hat{A}_s^{out}(t)=\hat X_s^{out}(t) + i \hat Y_s^{out}(t)
\EY
They are coupled to the intracavity amplitude $\hat{A}_s(t)$ by the boundary condition on the output mirror, which has the form
\BY
&&\hat A_{s}^{out}(t) = \sqrt{{\cal T}_s}\;\hat A_{s}(t)- \sqrt{{\cal R}_s}\;\hat A_{vac}(t),
\L{15}
\EY
where ${\cal T}_s$, ${\cal R}_s$ (${\cal T}_s+{\cal R}_s=1$) are the   reflection and transmission coefficients of the output mirror of the cavity (with ${\cal R}_s\approx1$).

The formal solutions of  Eqs.~(\ref{9}) read
\BY
 && \left(\begin{array}{c}\hat X_s(t,T)\\\hat Y_s(t,T)\end{array}\right)=
\int_{-\infty}^{T}d{T^\prime}
\left(\begin{array}{c}\hat{F}^\prime_{s}(t,
T^\prime)\\\hat{F}^{\prime\prime}_{s}(t, T^\prime)\end{array}\right)
e^{\dsp -\kappa_{\mp} (T-T^\prime)}.
\EY
The two-time representation was introduced only as an intermediate operation to obtain the expression of the intracavity pulses by solving a differential equation. We can now come back to the description of the successive pulses using:
\BY
 &&T\to nT_R,\qquad t\to t-nT_R,\qquad \hat X_s(t,T)\to\hat X_{s,n}(t-nT_R),\qquad \hat Y_s(t,T)\to\hat Y_{s,n}(t-nT_R).
\EY
We thereby get the intracavity quadratures which have to be used in Eq.~(\ref{15})
\BY
 && \left(\begin{array}{c}\hat X_{s,n}(t-nT_R)\\\hat Y_{s,n}(t-nT_R)\end{array}\right)=
\int_{-\infty}^{nT_R}d{T^\prime}  \left(\begin{array}{c}\hat{F}^\prime_{s}(t,
T^\prime)\\\hat{F}^{\prime\prime}_{s}(t, T^\prime)\end{array}\right) e^{\dsp -\kappa_{\mp} (nT_R-T^\prime)} .\L{18}
\EY
These expressions enable us to determine the correlation functions of the output field quadratures in terms of simple temporal integrals:
\begin{eqnarray}
&&\left(\begin{array}{c}\langle \hat X_{s,n}^{out}(t-nT_R)\; \hat
X_{s,n^\prime}^{out}(t^\prime-n^\prime T_R) \rangle\\\\\langle \hat Y_{s,n}^{out}(t-nT_R) \;\hat
Y_{s,n^\prime}^{out}(t^\prime-n^\prime T_R) \rangle\end{array}\right)
 =\nn\\&&\nn\\
 &&= \frac{1}{4} \delta_{nn^\prime} \delta(t-t^\prime)\pm \kappa_s T_R \frac{\mu(t-nT_R)/2}{1\mp\mu(t-nT_R)} e^{\displaystyle- \kappa_{\mp} (t) \; T_R|n-n^\prime|}
 \;\delta(t-t^\prime-(n-n^\prime) T_R).\L{19}
\end{eqnarray}
The first term is due to incoming vacuum field reflected from
coupling mirror of the cavity. The second term is related to the signal
field coming out of the cavity. We see that at discrete times $t\sim
nT_R$ the X-quadrature variance of a given pulse (when $n=n^\prime$)
is increased above the vacuum level, and the Y-quadrature is
squeezed. The presence of the delta-function in the expression is a
sign that the different temporal parts of an individual pulse are
not correlated, i.e., that the stretching/squeezing observed at
different times are independent. The importance of the effect is
defined by instantaneous value of the pumping amplitude, i.e., by
the pump parameter $\mu(t)$. It is actually small because it is
proportional to magnitude $\kappa_s T_R\ll1$.

Let us note that our model predicts quantum correlations between
different pulses (when $n\neq n^\prime$) for the X-quadrature of the
field, and anticorrelations for the Y-quadrature. The number of
significantly correlated successive pulses can be evaluated by the
factor in the exponential, which is proportional to $(\kappa_s
T_R)^{-1}$ and roughly equal to
cavity finesse at signal frequency. This result has the following
simple interpretation \cite{Reynaud}: the pump photons are
parametrically down-converted into pairs of correlated signal
photons. The photons of each pair may leave the cavity in different
pulses, during an overall time of the order of $\kappa_s^{-1}$,
giving rise to temporal correlations on the same range of time
difference. The delta-function shows that the correlations between
different pulses have a "local" character: they are effective only
when the time differences are a multiple of the period $T_R$.  The
equal time correlation of the signal field is obviously a
consequence of the thin nonlinear crystal approximation used in the
work.

To measure such quantum effects on the field quadratures, one must use a balanced homodyne detection technique, which will be considered in the next section.

\section{Balanced homodyne detection in the time domain\L{V}}
Let us now investigate the measurement of field quadratures, obtained by a balanced homodyne detection of the output signal field (see Fig.~2). In this case, the current operator has the well-known form
 \BY
 &&\hat i(t)=2\beta(t)\[ \cos\Phi\;\hat X_{s}^{out}(t)+\sin\Phi\;\hat Y_{s}^{out\dag}(t)\],\L{16}
\EY
where $\beta(t) e^{i \Phi}$ is the complex amplitude of the local oscillator. Choosing $\Phi=0$ and $\Phi=\pi/2$, one can follow both quadrature amplitudes in the form
\begin{eqnarray}
&&\hat i(t)=2 \sum_n\beta(t_n)\left(\begin{array}{c}\hat X_{s,n}^{out}(t_n)\\\\\hat
Y_{s,n}^{out}(t_n)\end{array}\right),\qquad t_n=t-nT_R\L{21}
\end{eqnarray}
We supposed here that the local
oscillator pulses are all identical, that they do not have phase
modulation and that their period is equal to the period of signal pulses
$T_R$. This implies that the envelope does not depend on pulse number $n$.
We have made the same assumptions for the SPOPO pumping field. The parameters that can be changed in the homodyne detection setup are the duration of the local oscillator pulses $\tau_{LO}$ as well as their delay $\Delta t$ relative to the signal pulses.

Substituting here the expression (\ref{19}), we derive the pair
correlation functions for the currents in the most general form
\BY
 &&\left(\begin{array}{c}\langle\hat i(t)\;\hat i(t^\prime)\rangle_X\\\\\langle\hat i(t)\;\hat
i(t^\prime)\rangle_Y\end{array}\right)=\nn\\
&&=\sum_n \beta^2(t-nT_R)\[\delta(t-t^\prime)\pm
  \kappa_s T_R\;\frac{2\mu(t-nT_R)}{1\mp\mu(t-nT_R)}e^{\ds-\kappa_\mp(t_n)\;|t-t^\prime|}\sum_{n^\prime}
 \delta\(t-t^\prime-nT_R+n^\prime T_R\)\]
.\L{22}
 \EY
Here as before we can conclude that in the time domain the quantum effects are of the order $\kappa_sT_R$ and are therefore very small. We will see in the next section that they appear more clearly on the noise spectra.

A real detector has a finite response time $T_D$, and time averages the photodetection signal, so that the observed photocurrent $\hat I(t)$ is given by
\begin{eqnarray}
&& \hat I(t) = \frac{1}{T_D}\int\limits_{t-T_D/2}^{t+T_D/2}dt\;\hat i(t).
\end{eqnarray}

Eq.~(\ref{22}) is valid when $T_D$ is much less than $T_R$. Let us now consider the case, important in practice, when $T_R\ll T_D\ll\kappa_s^{-1}$. This means that the pulse structure peculiar to the SPOPO is averaged. We expect therefore results looking like the c.w. regime. One gets in this case:
\BY
  &&\left(\begin{array}{c}\langle\hat I(t)\;\hat I(t^\prime)\rangle_X\\\\\langle\hat I(t)\;\hat I(t^\prime)\rangle_Y\end{array}
  \right)=\langle I\rangle\;
  \delta({t-t^\prime})\pm
\kappa_s
  A_\mp (t-t^\prime)
\L{crlt_i_x},
 \EY
 where
\BY
  &&\langle\hat I\rangle=
  \frac{1}{T_R}\int\limits_{-T_R/2}^{+T_R/2}d\tau\;\beta^2(\tau),\qquad A_\mp(t)=
    \frac{1}{T_R}\int\limits_{-T_R/2}^{+T_R/2}d\tau\;\beta^2(\tau)\frac{2\mu(\tau)}
    {1\mp\mu(\tau)}e^{\ds-\kappa_s\(1\mp\mu(\tau)\)|t|}
.\L{crlt_i_x_1}
 \EY
 By comparison with Eq.~(\ref{22}), wee see that only the equal time feature survives. Let us now choose the pulse of the local oscillator such that  $\beta$ as a function of $\tau$ is
 much narrower than $\mu(\tau)$; then it is possible to use the approximation $\beta^2(\tau)=\beta_0^2\;\delta(\tau)$, so that finally:

\BY
  &&\left(\begin{array}{c}\langle\hat I(t)\;\hat I(t^\prime)\rangle_X\\\\\langle\hat I(t)\;\hat I(t^\prime)\rangle_Y\end{array}
  \right)=\langle I\rangle\[
  \delta({t-t^\prime})\pm
\kappa_s
  \frac{2\mu(0)}
    {1\mp\mu(0)}e^{\ds-\kappa_s\(1\mp\mu(0)\)|t-t^\prime|}\],\qquad\langle\hat I\rangle=
 \frac{\beta_0^2 }{T_R}.
 \EY
Close to threshold ($1-\mu(0)\ll1$) we have:
\BY
  &&\langle\hat I(t)\;\hat I(t^\prime)\rangle_X/\langle I\rangle=
  \delta({t-t^\prime})+
\frac{\kappa_s}{1-\mu(0)}\;e^{\ds-\kappa_s(1-\mu(0))|t-t^\prime|}  ,\\
  &&{\langle\hat I(t)\;\hat I(t^\prime)\rangle_Y}/{\langle I\rangle}=
  \delta({t-t^\prime})-
\kappa_s\;e^{\ds-2\kappa_s|t-t^\prime|}  \L{}.
 \EY

\section{Quantum features on noise spectra\L{VI}}
In this section we determine the frequency spectrum of the quantum noise
on the $Y$-quadrature. This spectrum is defined as:
    \BY
    &&\(i^2\)_\Omega= \lim_{T\to\infty}\frac{1}{T}\int\limits_{-T/2}^{+T/2} dt\int\limits_{-T/2}^{+T/2}
    dt^\prime\;\langle\hat i(t)\;\hat i(t^\prime)\rangle_{Y} \;e^{\ds i\Omega (t-t^\prime)}\L{29}.
    \EY
In the case of $T_D\ll T_R$, and after substituting Eq.~(\ref{22})
into (\ref{29}), the spectrum reads:
    \BY
    &&\(i^2\)_\Omega = \frac{1}{T_R}\int\limits_{-T_R/2}^{+T_R/2}
    dt\;\beta^2(t)\(1-\sum_{m=0,1,2,\cdots}\frac{4\kappa_s^2\mu(t)}{\kappa_s^2\(1+\mu(t)\)^2+\(\Omega-2\pi
    m/T_R\)^2}\). \L{i2_w_full}
    \EY
This expression is valid for
arbitrary shapes of pump and local oscillator pulses $\mu(t)$ and $\beta(t)$.

In the case of very short local
oscillator pulses probing signal ones at their peaks we have the
simplified expression:
    \BY
    &&{\(i^2\)_\Omega/\langle I\rangle}= 1-\sum_{m=0,1,2,\cdots}
    \frac{4\kappa_s^2}{4\kappa_s^2+\(\Omega-2\pi m/T_R\)^2}. \L{i2_w}
    \EY
One can see that the quantum noise reduction takes place not only in
the vicinity of zero frequency, but also around all resonant
frequencies of the cavity $2 \pi m/T_R$, as known also for the c.w.
regime: indeed in the first experiment  on generation of squeezed
light based on four-wave mixing inside optical cavities
\cite{Slusher} the measured photon pairs were symmetrically shifted
with respect to the pump frequency by three cavity mode-spacing
frequencies. In \cite{Dunlop} and \cite{Senior} squeezing at
multiple longitudinal modes of a c.w. degenerate OPO was
investigated theoretically and observed experimentally.

The actual frequency response of detectors can be taken into account
by multiplying the spectrum (\ref{i2_w}) by the detector frequency
response function \cite{Goodman}. Obviously a high bandwidth
photodetection set-up must be used to observe the noise reduction
around multiple resonant frequencies. However if $T_R\ll
T_D\ll\kappa_s^{-1}$, then in the spectrum only one quantum feature
survives around the zero frequency and its spectrum is given by the
well-known formula
    \begin{eqnarray}
    && {\(I^2\)_\Omega/\langle I\rangle}= 1-\frac{4\kappa_s^2}{4\kappa_s^2 +\Omega^2}.
    \end{eqnarray}

Turning back to the general expression (\ref{i2_w_full}),  one sees that it is nothing else than the time averaged noise spectrum of the signal field quadrature of a c.w. OPO pumped by a c.w. field of instantaneous value $\mu(t)$, weighted by the intensity of the local oscillator pulse. This is a consequence of the fact that different temporal parts of individual pulses of SPOPO are not correlated in our simple model and their
nonclassical properties are defined by the instantaneous pump power. Consequently, the detected quantum noise is sensitive to the temporal properties
of the local oscillator pulses, particularly to their duration and
to the delay relative to signal pulses (see Figures \ref{fig:i^2_LO} (a) and (b) respectively). As expected the model predicts maximal noise reduction for very
short local oscillator pulses ideally synchronized with the signal maxima.

\section{Conclusion\L{VII}}
In this paper, we have investigated the non-classical properties of a SPOPO operated below threshold using a time-domain approach,  in the simplified case when it is possible to neglect the dispersion phenomenon. We have established differential equations for the intracavity field operator amplitude and on this
basis investigated the quantum statistical properties of the signal field.  As expected the typical
region of correlation is the same as for the c.w. regime. However there are supplementary details related to the pulse field structure, namely the fact that different times inside the limits of the same pulse turn out to be uncorrelated, whereas small but nonzero inter-pulse quantum correlations occur between fluctuations at times similarly placed in the different pulses.

We have shown that the "instantaneous" homodyne signal (i.e. using a local oscillator of very short duration) turns out to be perfectly squeezed when approaching the oscillation threshold from below and at noise frequencies
 $2\pi m /T_R$ ($m=0,1,2,\cdots$), just like in the c.w. case. From a practical point of view the major advantage of the SPOPO is its very low oscillation threshold in terms of mean pump power. This implies that one can use a moderately resonant
cavity, with a high escape efficiency, and still have a threshold that can be reached using available mode-locked lasers with average powers in the 100mW range.

Let us mention finally that the Heisenberg-Langevin equations that
we have established can be modified straightforwardly to take into
account experimental effects such as phase modulation and
carrier-envelope phase shift of pump pulses, cavity detuning, and
singly resonant operation.

\section{Acknowledgements\L{VIII}}

We acknowledge helpful discussions with G. Patera. The study was performed within the framework of
the Russian-French Cooperation Program "Lasers and Advanced Optical Information Technologies" and
the European Project HIDEAS (grant No. 221906). It was also supported by RFBR (No. 08-02-92504, and
No. 08-02-00771). VA acknowledges financial support by French government grant (No.
0185-RUS-B09-0529).

\appendix

\section{Some quantitative estimations}\L{estimations}
In this {appendix} we estimate the effects that occur in the
propagation of femtosecond pump and signal pulses through a
dispersive $\chi^{(2)}$ nonlinear crystal. These processes are the
reshaping and chirping of pulses due to group velocity dispersion of
the crystal and the pulse temporal walk-off due to mismatch of group
velocities. The efficiency of each process can be described by a
characteristic propagation distance \cite{Becker}.
The group velocity mismatch of the pump and signal fields ($\upsilon_p$ and $\upsilon_s$ respectively)
is characterized by the propagation distance $L^{GV}$ necessary for a signal pulse to walk through the pump one
\begin{eqnarray}
&& L^{GV}   = \frac{\tau_p}{|1/\upsilon_s-1/\upsilon_p|}, \L{GV}
\end{eqnarray}
where $\tau_p$ is the duration of pump pulses. The linear dispersion of the
crystal is characterized by the distance $L^{D}$ over which the initial duration of Gaussian pulse
$\tau_{p,s}$ will increase by a factor of $\sqrt{2}$
\begin{eqnarray}
&& L_{p,s}^{D}  = \frac{\tau_{p,s}^2}{2\beta_{p,s}} \L{D}
\end{eqnarray}
The efficiency of parametric interaction is characterized by the distance $L^{NL}$
\begin{eqnarray}
&& L^{NL}   = \frac{1}{2\sigma |A_{0}^{peak}|} \L{NL}
\end{eqnarray}
where $A_0^{peak}$ - pump peak amplitude defined by external pumping as well as duration of pump
pulses.
For straightforward parametric amplification with a pump peak amplitude the signal wave amplitude
grow by a factor of $e$ over this propagation distance.

In Table 1 we present characteristic distances evaluated for the following experimental parameters
considered in \cite{Patera}: pump wavelength $\lambda_p = 0.4 \mu m$, duration of pump pulses -
$\tau_p = 100 fs$, c.w. threshold pump power inside the cavity - $P_{th} = 25 mW$; crystal length -
$l = 0.1 mm$.
We have also supposed following duration of signal pulse $\tau_s = 100 fs$ in order to estimate
$L_{s}^{D}$ using (\ref{D}).

\begin{center}
\begin{tabular}{|@{\quad}c@{\quad}|@{\quad}c@{\quad}|@{\quad}c@{\quad}|@{\quad}c@{\quad}|@{\quad}c@{\quad}|}
\hline \hline
crystal & $L^{GV}$, mm & $L_p^{D}$, mm & $L_s^{D}$, mm ($\tau_s=100$ fs) & $L^{NL}$, mm\\
\hline
BBO & $\sim (0.25\div1)$ & $\sim (10\div20)$ & $\sim (20\div30)$ & $\sim4.7$\\
\hline
KNbO$_3$ & $\sim (0.1\div0.75)$ & $\sim (10\div50)$ & $\sim (20\div40)$ & $\sim1.6$\\
\hline
\end{tabular}
\end{center}
\begin{center}
Table 1: Estimations of characteristic distances
\end{center}
Thus data of the Table 1 show that {for used} experimental parameters the following relation holds
\begin{eqnarray}
&& l \leq L^{GV} < L^{NL} < L^{D} \L{ranging}
\end{eqnarray}
Therefore, in the cases we consider here all the discussed processes
can be considered as small perturbations of free propagation of pump
and signal fields inside the crystal (both BBO and KNbO$_3$).
More precisely the dispersive reshaping and chirping of envelopes of fields are the weakest processes.

Expression (\ref{D}) also defines characteristic time for the crystal of given length $l$
\begin{eqnarray}
&& \tau_s^D  = \sqrt{2 \beta_s l} \L{tD}
\end{eqnarray}
Thus, neglecting crystal dispersion we can
correctly describe the field properties only at time scales larger than $\tau_s^D$. For the present experimental parameters this time is on the order of 6 fs.

\section{Derivation of the time-domain equations for the intracavity pulsed field} \L{HLE}

Let us write the pump and signal fields propagating through the crystal in terms  of slowly varying
complex envelopes $\hat A_r(z,t)$ {introduced} by means of the definition
\BY
&& \hat E_r(z,t)=i \left(\frac{\hbar \omega_r}{2 n_r \varepsilon_0 c S} \right)^{1/2}e^{-i\omega_r t+ik_rz}\hat
A_r(z,t),\qquad r=s,p,\qquad(\omega_p=2\omega_s).\L{A1}
\EY
Here $k_r=k_r(\omega_r)$ are the pump and signal wave vectors determined by linear dispersion of the crystal; {$n_r =
n_r(\omega_r)$ - the indexes of refraction of the crystal.}

Collinear propagation of pump and signal fields inside $\chi^{(2)}$
nonlinear crystal with dispersion is described by following two
coupled traveling-wave equations written for {operators of complex
envelopes} of fields \cite{Becker}

\begin{eqnarray}
&& \left( \frac{\partial }{\partial z}+\frac{1}{\upsilon_{p}}\frac{\partial }{\partial t}+i \frac{\beta_{p}}{2}
\frac{\partial^{2} }{\partial t^2} \right) \hat A_{p}(z,t) = - \sigma \hat A_s^2(z,t)  \;e^{-i \Delta\! k z}, \L{A2}\\
&& \left( \frac{\partial }{\partial z}+\frac{1}{\upsilon_{s}}\frac{\partial }{\partial t}+i \frac{\beta_{s}}{2}
\frac{\partial^{2} }{\partial t^2} \right) \hat A_s(z,t) =  2\sigma \hat A_{p}(z,t)\hat A_s^{\dag}(z,t) \;e^{i
\Delta \!k z}\L{A3}
\end{eqnarray}
{Here $\sigma$ is the coupling constant, proportional to the nonlinear susceptibility of the crystal $\chi^{(2)}$ }
\begin{eqnarray}
&& \sigma = \frac{\chi^{(2)}}{c} \; \sqrt{ \frac{\hbar \omega_p \; \omega_s^2}{2 \epsilon_0 c S n_p n_s^2}   }.
\end{eqnarray}
The inverse  group velocities and  the dispersions of group velocities are respectively
\BY
&& \upsilon_{r}^{-1}=  \frac{\displaystyle \partial k_{r}(\omega)}{\displaystyle \partial \omega} |_{\omega=\omega_r
},\qquad\beta_{r}=\frac{\displaystyle \partial^2 k_{r}(\omega)}{\displaystyle \partial
\omega^2}|_{\omega=\omega_r}.\L{A5}
\EY
Exponents on the right hand sides of equations describe phase mismatch of pump and signal carrying waves along their
propagation inside the crystal
\BY
&&\Delta \!k = k_p(\omega_p)-2k_s(\omega_s).\L{A6}
\EY
In this paper we consider degenerate regime of SPOPO operation,
$\Delta\! k = 0$.

These equations assume an instantaneous nonlinear response of the medium and use the approximation  of
slowly varying envelopes. Also linear dispersion of the crystal is considered up to the second
order in the perturbative expansion around the pump and signal carrying frequencies \cite{Chirkin}.
Under these approximations equations do not describe arbitrarily fast dynamics of envelopes of
fields, but they are quite reasonable for pulses with duration of tens of femtoseconds.

Let us assume that $\beta_r=0$, meaning that the parametric crystal is thin enough so that the dispersive reshaping  and chirping of envelopes of fields along their propagation inside the crystal are negligible. Under this approximation second-order time derivatives in equations (\ref{A2})
and (\ref{A3}) are dropped. We also assume the parametric interaction is very weak in a single pass through the
nonlinear crystal, so that we can use in the right hand sides of the equations the amplitudes of the free propagating fields. In addition, we will assume for simplicity that the group velocities of the pump and signal fields are equal: $\upsilon_p=\upsilon_s=\upsilon$.
The traveling-wave equations take now the simplified form:
\begin{eqnarray}
&& \left( \frac{\partial }{\partial z}+\frac{1}{\upsilon}\frac{\partial }{\partial
t} \right) \hat A_{p}(z,t) = -
\sigma \hat A_s^2(0,t-z/\upsilon), \L{A7}\\
&& \left( \frac{\partial }{\partial z}+\frac{1}{\upsilon}\frac{\partial }{\partial t} \right)
\hat A_s(z,t) = 2
\sigma \hat A_{p}(0,t-z/\upsilon)\hat A_s^{\dag}(0,t-z/\upsilon) \L{A8}
\end{eqnarray}
The solutions of these equations can be easily obtained by replacing the time $t$ by its reduced
value $\eta = t - z/\upsilon$ and
integrating obtained equations over the crystal length. The solutions read
\begin{eqnarray}
&& \hat A_{p}(l,t) = \hat A_{p}(0,t-l/\upsilon) - \sigma l \;{\hat A_s^2(0,t-l/\upsilon)} \L{OPA_p_1},\\
&& \hat A_s(l,t) = \hat A_s(0,t-l/\upsilon) + 2\sigma l \;{ \hat A_{p}(0,t-l/\upsilon)} \hat
A_s^{\dag}(0,t-l/\upsilon) \L{OPA_s_1}
\end{eqnarray}

Now let this thin parametric crystal be placed inside the high-finesse cavity just near the input
mirror (see Fig. 1). Putting the
coordinate axis $z$ parallel to the optical axis of the cavity one can couple by  the following
relation the intracavity field amplitudes $\hat A_r(L,t)$ and $\hat A_r(0,t)$ before and after the
coupling mirror respectively with the amplitude  $\hat A^{in}_{r}(0,t)$ of the external field
entering the cavity
\BY
&&\hat A_r(0,t )=\sqrt{{\cal R}_r}\;\hat A_{r}( L,t)+\sqrt{{\cal T}_r}\;\hat A_{r}^{in}(0,t),
\L{mirror_1}
\EY
{where ${\cal R}_r$ and ${\cal T}_r$ ($r=p,s$) are reflection and transmission  coefficients of the
mirror.}
The slow amplitudes of the input fields defined in the same manner as intracavity ones obey
following commutation relations \cite{Kolobov_99}
\begin{eqnarray}
&& [\hat{A}_r^{in}(z,t), \hat{A}^{in\dag}_{r}(z,t^\prime)] =   \delta(t-t^\prime), \quad [\hat{A}_r^{in}(z,t),
\hat{A}_{r}^{in}(z,t^\prime)] = 0. \L{3}
\end{eqnarray}

Using solutions (\ref{OPA_p_1})-(\ref{OPA_s_1}) obtained under thin crystal approximation   and
assumption of equal group velocities it is straightforward to write following the relations that couple
the amplitudes of fields before and after one round-trip inside the SPOPO cavity
\begin{eqnarray}
&& \hat A_{p}(L,t) = \(\hat A_{p}(0, t-T_{R}) - \sigma l \hat A_s^{2}(0, t-T_{R})\) \; e^{i \omega_p T_{ph,p}} \L{rnd_trip_p} ,\\
&& \hat A_{s}(L,t) = \(\hat A_{s}(0, t-T_{R}) + 2 \sigma l \hat A_p(0, t-{T_R}) \hat A_s^{\dag}(0,
t-T_{R})\)\; e^{i \omega_s  T_{ph,s}} \L{rnd_trip_s}
\end{eqnarray}
Here $T_{R}$ - round-trip time of pump and signal pulses inside the cavity; $T_{ph,p}$ and $T_{ph,s}$ - round-trip times of pump and signal carrier waves.
We assume that the pump and signal carriers are resonant for the cavity so that conditions $\omega_p T_{ph,p} = 2\pi m$ and $\omega_s \; T_{ph,s} = 2\pi n$ hold.

Representing continuous slowly varying envelope of the signal field
as the following piecewise function:
\BY
&& \hat A_s( 0,t)=\sum_{n}\hat A_{s,n}(t-nT_R) \L{A16}
\EY
and combining (\ref{mirror_1}) with (\ref{rnd_trip_s}) one obtains
\BY
&&\hat A_{s, n}(t-nT_R)=\sqrt{{\cal R}_s}\(\hat A_{s, n-1}(t-(n-1)T_R) + 2\sigma l \;{ \hat A_{p, n-1}(t-(n-1)T_R)}
\hat A_{s, n-1}^{\dag}(t-(n-1)T_R)\)+\nn\\
&&+\sqrt{{\cal T}_s}\;\hat A_{s,n}^{in}(t-nT_R),\L{A17}
\EY
where the time $t$ is treated as the time deviation from the pulse center. We see that there are actually
two time arguments, because the index $n$ informs us about exchanges from pulse to pulse. Strictly
speaking this second argument is discrete with distance between adjoining point equal to $T_R$.
However assuming that the pulse envelope does not change essentially after one round
trip inside the cavity one can neglect the time interval $T_R$ in comparison with other typical. This enables us to consider the time variable $T$ as continuous. Such a two-time approach has been used in Ref.~\cite{Haus} for description of pulsed lasers. We can then make the following replacements in Eq.~(\ref{A17})%
\BY
 &&\hat A_{r,n}(t-nT_R)\to
 \hat A_{r}(t, T)\;(r=p,s),\qquad\frac{\hat A_{s,n}(t-nT_R)-\hat A_{s,n-1}(t-(n-1)T_R)}{T_R}\to
 \frac{\partial\hat A_{s}(t, T)}{\partial T}\L{A18}
\EY
We can finally use the same procedure to get the equation for the pump pulse amplitude. We have then justified in this appendix the introduction of equations ~(\ref{HL_p}) and (\ref{HL_s}) of the present paper.

\newpage

\begin{figure}
\centering
\includegraphics[width=15cm]{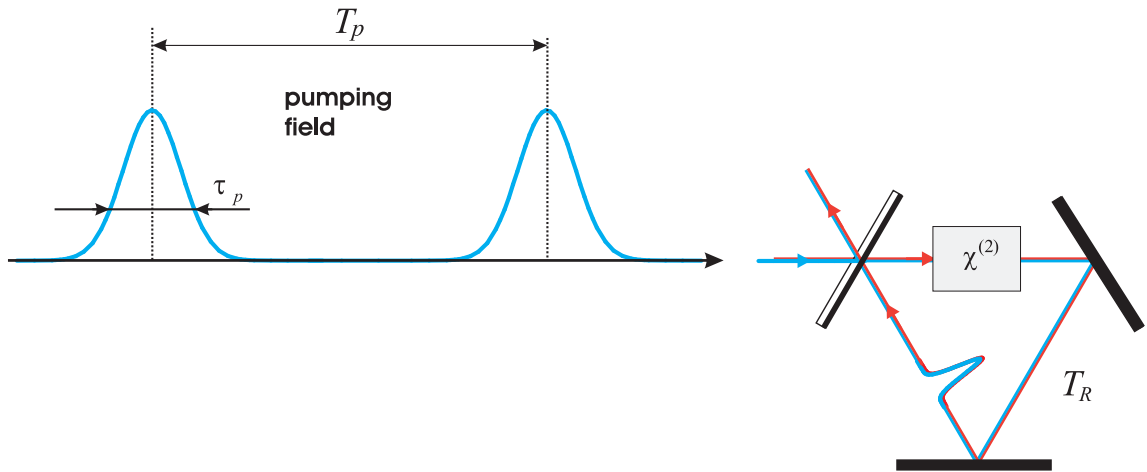}
\caption{Synchronously pumped OPO}
\label{fig:SPOPO}
\end{figure}

\begin{figure}
\centering
\includegraphics[width=15cm]{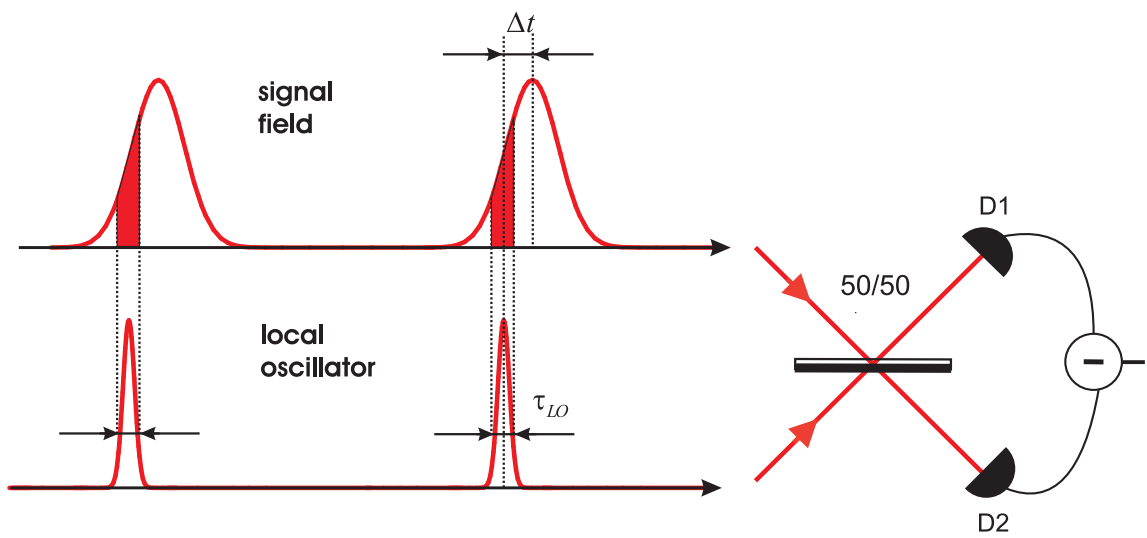}
\caption{Balanced homodyne detection}
\label{fig:BHD}
\end{figure}

 \begin{figure*}
  \centerline{
    \mbox{\includegraphics[height=8cm]{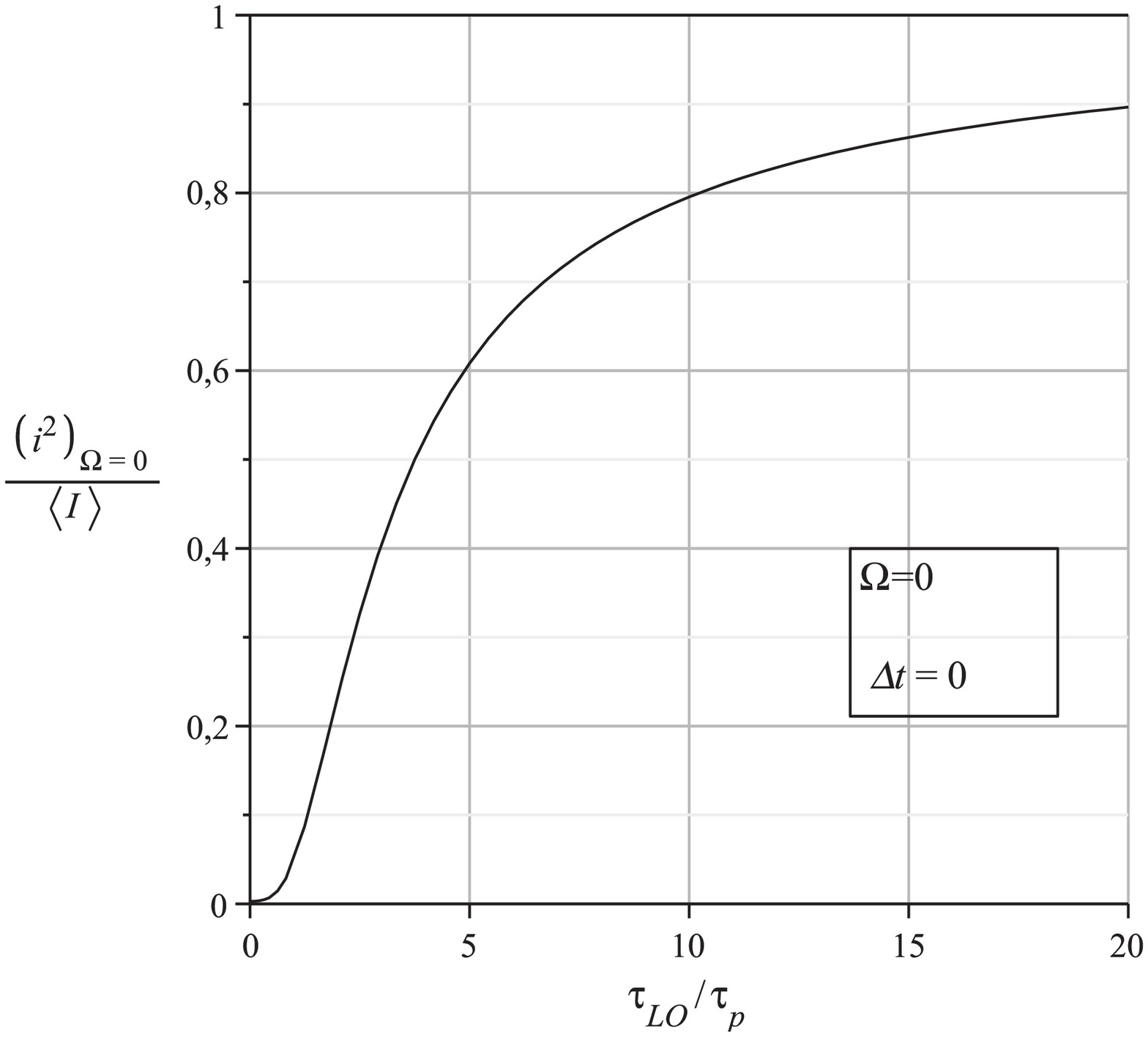}}
    \mbox{\includegraphics[width=8cm]{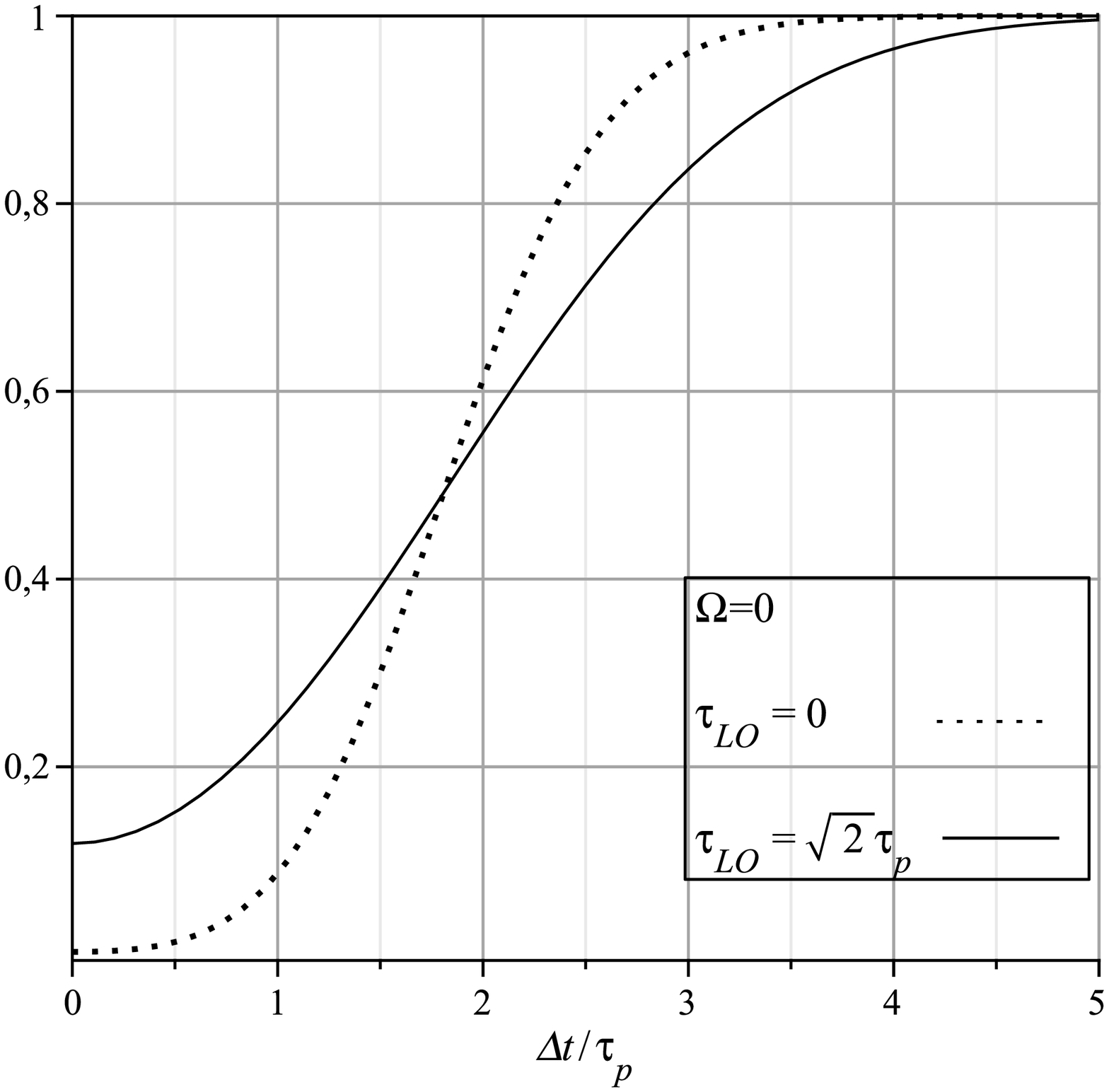}}
  }
  \caption{Quantum noise of the squeezed quadrature normalized to the shot noise at zero frequency
  as a function of duration of local oscillator (LO) pulses (a) and their delay relative to signal ones (b). Both times normalized to pump pulses duration.  LO pulses and pump ones are Gaussian.
  At the Fig. (a) LO pulses are ideally synchronized with signal $\Delta t = 0$;
  at the Fig. (b) two fixed durations of the pulses are considered. In both cases $\kappa_s T_R =0.1$ and $\mu(0) = 0.9$.}
  \label{fig:i^2_LO}
  \end{figure*}

\end{document}